\documentclass[11pt,oneside,a4paper,onecolumn]{elsarticle}

\usepackage{times}
\usepackage{fullpage}

\usepackage{gensymb}
\usepackage{alltt}
\usepackage{upquote}
\usepackage{float}
\usepackage{array}
\usepackage{pgfplots}
\usetikzlibrary{positioning}
\usepackage{graphicx}
\usepackage[toc,page]{appendix}
\usepackage[utf8]{inputenc}
\usepackage{amsmath,mathtools}
\usepackage{amsthm}
\usepackage{amsfonts}
\usepackage{amssymb}
\usepackage{enumitem}
\usepackage{caption}
\usepackage{subcaption}
\usepackage{xcolor}
\usepackage{scalerel}
\usepackage{algorithm}
\usepackage{algorithmic}

\usepackage{multicol}
\usepackage{multirow}
\usepackage{url}
\usepackage{booktabs}
\usepackage{nicefrac}

\usepackage{nowidow}

\usepackage{soul}
\usepackage{slashed}
\usetikzlibrary{patterns}
\usetikzlibrary{shapes.arrows}
\usetikzlibrary{fillbetween,backgrounds}
\usepackage{placeins}
\pgfplotsset{compat=newest}
\usepackage{diagbox}
\usetikzlibrary{arrows.meta}

\pgfdeclarelayer{nodelayer}
\pgfdeclarelayer{edgelayer}

\pgfsetlayers{background,nodelayer,edgelayer,main}

\usepackage[final]{hyperref} 
\hypersetup{
colorlinks=true,       
linkcolor=blue,        
citecolor=blue,        
filecolor=magenta,     
urlcolor=blue         
}
\usepackage[capitalize]{cleveref}
\creflabelformat{equation}{#2\textup{#1}#3}

\newcolumntype{M}[1]{>{\centering\arraybackslash}m{#1}}
\newcolumntype{N}{@{}m{0pt}@{}}
\newtheorem{remark}{Remark}

\tikzset{>=latex}
\def\@author#1{\g@addto@macro\elsauthors{\normalsize%
	\def\baselinestretch{1}%
	\upshape\authorsep#1\unskip\textsuperscript{%
		\ifx\@fnmark\@empty\else\unskip\sep\@fnmark\let\sep=,\fi
		\ifx\@corref\@empty\else\unskip\sep\@corref\let\sep=,\fi
	}%
	\def\authorsep{\unskip,\space}%
	\global\let\@fnmark\@empty
	\global\let\@corref\@empty
	\global\let\sep\@empty}%
\@eadauthor={#1}
}

\graphicspath{{Images/}}

\begin{document}
\begin{frontmatter}
	\title{A Mesh-Adaptive Hypergraph Neural Network for Unsteady Flow Around Oscillating and Rotating Structures}
	\author[ubc]{Rui Gao}
	\ead{garrygao@student.ubc.ca}
	
	\author[ucb]{Zhi Cheng}
	\ead{chengzhi@berkeley.edu}
	
	\author[ubc]{Rajeev K. Jaiman\corref{cor1}}
	\ead{rjaiman@mech.ubc.ca}
	\cortext[cor1]{Corresponding author}
	\address[ubc]{Department of Mechanical Engineering, The University of British Columbia, Vancouver, BC V6T 1Z4}
	\address[ucb]{Department of Mechanical Engineering, University of California, Berkeley, Berkeley, CA 94720}
	
	\begin{abstract}
		Graph neural networks, recently introduced into the field of fluid flow surrogate modeling, have been successfully applied to model the temporal evolution of various fluid flow systems. Existing applications, however, are mostly restricted to cases where the domain is time-invariant. The present work extends the application of graph neural network-based modeling to fluid flow around structures rotating with respect to a certain axis. Specifically, we propose to apply a graph neural network-based surrogate model with part of the mesh/graph co-rotating with the structure and part of the mesh/graph static. A single layer of interface cells are constructed at the interface between the two parts and are allowed to distort and adapt, which helps in circumventing the difficulty of interpolating information encoded by the neural network at every graph neural network layer. Dedicated reconstruction and re-projection schemes are designed to counter the error caused by the distortion and connectivity change of the interface cells.
		The effectiveness of our proposed framework is examined on two test cases: (i) fluid flow around a 2D oscillating airfoil, and (ii) fluid flow past a 3D rotating cube. Our results show that the model achieves stable rollout predictions over hundreds or even a thousand time steps. We further demonstrate that one could enforce accurate, error-bounded prediction results by incorporating the measurements from sparse pressure sensors. In addition to the accurate flow field predictions, the lift and drag force predictions closely match with the computational fluid dynamics calculations, highlighting the potential of the framework for modeling fluid flow around rotating structures, and paving the path towards a graph neural network-based surrogate model for more complex scenarios like flow around marine propellers.
		
		\smallskip
		\smallskip
		
	\end{abstract}
\end{frontmatter}

\section{Introduction}
\label{sec:intro}
Rotating structures, including marine and aeronautic propellers, centrifugal pumps and blowers, stirring mixers, wind and gas turbines, etc., comprise an important part of modern machinery. Most of these rotating structures are immersed in fluids, usually air or water, and the fluid flow around such rotating structures is frequently of interest due to their profound impact on the performance and lifespan of machines that include these rotating structures. To improve the understanding of the fluid flow around rotating structures and to facilitate design and control efforts, computational fluid dynamics (CFD) techniques are usually adopted to generate simulations for such fluid systems \cite{keck2008thirty,pinto2017computational,cheng2022unified,posa2022hydroacoustic}. However, CFD simulations can be computationally heavy, especially when simulating large 3D cases. This limitation leads to challenges in the application of CFD simulations in the iterative design process and real-time model-based flow control. An increasing interest has therefore arisen in the development of cheaper surrogate models that could be adopted for the aforementioned purposes. 

Recent rapid development and application of deep learning methods has inspired a series of applications of these models in fluid flow modeling, including the modeling of fluid flow around rotating structures. Existing work includes predicting or fitting fluid field within a pump \cite{li2024deep}, around a propeller \cite{li2024fast,hou2024reconstruction}, within a stirring mixer \cite{gao2022simulation,zhang2024rotating}, around a spinning sphere \cite{liu2024parameterized}, within a co-rotating disk cavity \cite{zhang2024performance}, etc. These works typically adopt either convolutional neural networks or multilayer perceptrons with specially-designed features or training losses. The target in most of these works, however, is to obtain a prediction of static or time-averaged flow field rather than the spatio-temporal evolution of the flow. One significant exception is the recent work by Li et al. \cite{li2024fast} which features time series prediction of the wake after a propeller with a merged convolution-transformer network, under a series of limitations: Only the flow in the wake region is predicted, only a 2D slice out of the 3D flow field is predicted, and the authors only report predictions up to 15 time steps -- a rather short rollout compared with standard results in 2D flow scenarios. 

Recently, graph neural networks (GNNs) have been introduced to the modeling of fluid flow. Being able to work with graphs transformed from unstructured mesh, one could easily adopt a body-fitting mesh that is only refined in regions of interest, leading to significant reduction in the mesh size compared to the case with convolutional neural networks while maintaining a satisfactory resolution. This unique advantage has inspired the recent integration of graph neural networks into various fluid flow modeling scenarios. Applications include, but are not limited to, flow around bluff or streamlined bodies \cite{pfaff2020learning,lino2022multi,lino2024se,fortunato2022multiscale,cao2022bi,gao2024finite}, fluid-structure interactions \cite{gao2024predicting}, particulate suspension \cite{ma2022fast}, reacting flow \cite{xu2021conditionally}, cavitating flow \cite{gao2024towards}, shock waves \cite{li2024physics}, and environmental flow above \cite{kazadi2024floodgnn} or below \cite{tang2024graph} the surface of the Earth. For a more comprehensive review of recent advances in the field, the reader is referred to Zhao et al. \cite{zhao2024review}. Considering the variety of reported applications of graph neural networks, it is reasonable to hypothesize that it will also serve as a proper architecture for modeling fluid flow around rotating structures. 
Li et al. \cite{li2023uncertainty} recently applied a graph neural network to model the flow around a radial turbine. However, similar to previous works using other network architectures, their focus was limited to predicting time-averaged flow fields, without addressing the spatio-temporal dynamics of unsteady flow around rotating structures. They also chose to only model only a single blade of the turbine, which relies on the implicit assumption that the flow is cyclic periodic that can be true for the time-averaged flow field but not for instantaneous flow fields.

In this work, we aim to bridge this gap by developing a graph neural network framework capable of modeling the full spatio-temporal evolution of fluid flow around rotating bodies. Specifically, we partition the domain into two sub-domains, one co-rotating with the solid body and one static. A single layer of cells is constructed at the interface between the two sub-domains and is allowed to adapt over time. 
A graph neural network is employed to predict the evolution of the system states over each time step, and a series of reconstruction and re-projection steps are designed and implemented to account for the mesh distortion and adaptation between the time steps. Predictions from the framework are enforced to be equivariant over the rotation of the system domain via appropriate input and output feature transformations.

The proposed graph neural network framework will be assessed in two test cases, namely the flow around a 2D oscillating airfoil and the flow around a 3D rotating cube. We will demonstrate that the network is able to provide stable long-term rollout predictions for both cases. For the 2D case, we additionally demonstrate that one could enforce accurate, error-bounded prediction results by integrating the measurements from sparse pressure sensors mounted near the surface of the solid body.
These results highlight the ability of graph-based surrogate modeling for fluid flow around rotating bodies and provide a foundation for extending the method to more complex systems such as propellers, turbines, and other real-world applications.

This paper is organized as follows. Section \ref{sec:methodology} will cover the necessary background and methodology, including the definition of the problem along with its domain, the spatio-temporal discretization of the system, the setup of the GNN-based framework, as well as the mesh construction and adaptation algorithm. In the following Section \ref{sec:framework} we assemble these components into a complete framework, suitable for modeling fluid flow around rotating structures. We lay out the setup of two test cases used to demonstrate the framework as well as the implementation and training details for the framework in Sec. \ref{sec:setup}. The results of the two test cases will be discussed and compared with the corresponding CFD results in Sec. \ref{sec:results}. We will conclude the work in Section \ref{sec:conclusion}.

\section{Methodology}
\label{sec:methodology}
\subsection{Fluid flow around rotating structure, spatial and temporal discretization}
\label{sec:discretization}
\begin{figure}
	\centering
	\includegraphics[]{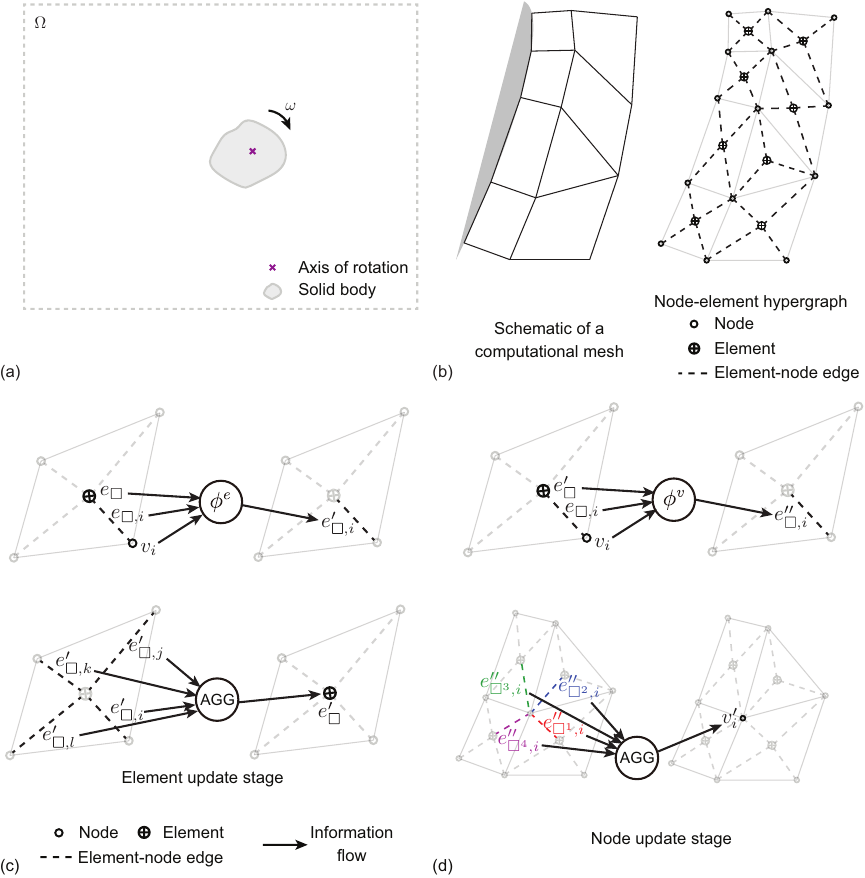}
	\caption{(a) Schematic of the system domain $\Omega$ which contains a rotating solid body, (b) conversion from a computational mesh to a node-element hypergraph, and (c) schematic of the element and node update stages within each hypergraph message-passing layer in $\phi$-GNN.}
	\label{fig:dgmp}
\end{figure}

Consider a domain $\Omega$ in which a rigid body immersed in fluid flow is rotating around a fixed axis, shown in Fig. \ref{fig:dgmp}a. We seek an efficient surrogate model $\widehat{G}$ for the spatio-temporal evolution of such a flow system, the governing equation of which can be written into an abstract dynamical form as
\begin{equation}
	\label{eq:state}
	\frac{d\boldsymbol{q}}{dt}=\tilde{F}(\boldsymbol{q}),
\end{equation}
with $\tilde{F}$ being the governing model for the system. Applying forward Euler temporal discretization with a fixed time step gives
\begin{equation}
	\label{eq:forwardEuler}
	\boldsymbol{q}^{t_{n+1}}-\boldsymbol{q}^{t_n}=\delta\boldsymbol{q}=F(\boldsymbol{q}(t_n)),
\end{equation}
where $\boldsymbol{q}^{t_n}=(\boldsymbol{u}^{t_n},p^{t_n})$ denotes the system state variables at the $n$-th time step, and $F$ is the state update function after temporal discretization. Assuming that the mesh is invariant within each time step, we further discretize Eq. \ref{eq:forwardEuler} spatially with the mesh at each time step $t_n$, leading to 
\begin{equation}
	\label{eq:mesh}
	\boldsymbol{Q}^{t_{n+1}}-\boldsymbol{Q}^{t_n}=\delta\boldsymbol{Q}=\boldsymbol{F}(\boldsymbol{Q}^{t_n}),
\end{equation}
where $\boldsymbol{Q}^{t_n}$ is the matrix containing system state variables at the $n$-th time step, and $\boldsymbol{F}$ is the function to update the state variables over each time step. 

As mentioned in the Introduction, one could easily convert a mesh to a graph. If one performs such conversion from a mesh to a hypergraph $\mathcal{G}=(\mathcal{V},\mathcal{E}_\square,\mathcal{E}_{v})$ containing the set of all nodes $\mathcal{V}$, the set of all elements $\mathcal{E}_\square$, as well as the set of all element-node edges $\mathcal{E}_{v}$ (cf. Fig. \ref{fig:dgmp}b), then one could also transform the system states $\boldsymbol{Q}$ into a series of features and attach them to the hypergraph as feature vectors. We delay the discussion of these transformations to Sec. \ref{sec:featvec}, and for now simply assume that the system states at each time step $t_n$ are transformed to node feature vector $\boldsymbol{v}_i^{t_n}$ for each node $i$, element feature vector $\boldsymbol{e}_{\square}^{t_n}$ for each element $\square$, as well as element-node edge feature vector $\boldsymbol{e}_{\square,i}^{t_n}$ for each element-node edge connecting node $i$ and element $\square$. The discretized time stepping (Eq. \ref{eq:mesh}) can then be rewritten as
\begin{equation}
	\label{eq:graphstepfull}
	\boldsymbol{v}_i^{t_{n+1}}-\boldsymbol{v}_i^{t_{n}},\boldsymbol{e}_{\square}^{t_{n+1}}-\boldsymbol{e}_{\square}^{t_n},\boldsymbol{e}_{\square,i}^{t_{n+1}}-\boldsymbol{e}_{\square,i}^{t_n}=\boldsymbol{G}(\boldsymbol{v}_i^{t_{n}},\boldsymbol{e}_{\square}^{t_n},\boldsymbol{e}_{\square,i}^{t_n}),
\end{equation}
in which $\boldsymbol{G}$ is a single or a set of functions that update the hypergraph features over each time step. In this work, we design the feature transformations in a way such that all the features are on element-node edges (cf. Sec. \ref{sec:featvec}), and then Eq. \ref{eq:graphstepfull} can be simplified as
\begin{equation}
	\label{eq:graphstep}
	\boldsymbol{e}_{\square,i}^{t_{n+1}}-\boldsymbol{e}_{\square,i}^{t_n}=\boldsymbol{G}(\boldsymbol{e}_{\square,i}^{t_n}).
\end{equation}
A hypergraph neural network model $\widehat{\boldsymbol{G}}$ can be constructed as the surrogate for $\boldsymbol{G}$ in Eq. \ref{eq:graphstep}. We briefly cover the architecture of the network architecture used in this work in the subsequent Sec. \ref{sec:model}.

\subsection{Network architecture}
\label{sec:model}
We adopt the recently proposed finite element-inspired $\phi$-GNN \cite{gao2024finite} in the current work. Operating on a hypergraph discussed in the previous Sec. \ref{sec:discretization}, the network mimics the local stiffness matrix calculation process in the finite element method, offering state-of-the-art performance along with explainability. Following the encode-propagate-decode fashion, the network contains a series of hypergraph message-passing layers between the encoding functions at the start and a decoding function at the end. With input node feature vectors $\boldsymbol{v}_i$, element feature vectors $\boldsymbol{e}_{\square}$, and element-node edge feature vectors $\boldsymbol{e}_{\square,i}$, different encoding functions $g^v$, $g^e$, and $g^{ev}$ are applied to respective feature vectors:
\begin{subequations}
	\begin{equation}
		\boldsymbol{v}_i\leftarrow g^v(\boldsymbol{v}_i),
	\end{equation}
	\begin{equation}
		\boldsymbol{e}_\square\leftarrow g^e(\boldsymbol{e}_\square),
	\end{equation}
	\begin{equation}
		\boldsymbol{e}_{\square,i}\leftarrow g^{ev}(\boldsymbol{e}_{\square,i}).
	\end{equation}
\end{subequations}
The left arrow $\leftarrow$ denotes an update of the parameter on its left-hand side by the value on its right-hand side. Each of the subsequent message-passing layers $\widehat{G}_m$ contain an element update stage and a node update stage illustrated in Fig. \ref{fig:dgmp}c and \ref{fig:dgmp}d. Assuming a hypergraph converted from a 2D quadrilateral mesh, for each layer $m$, the two update stages can be written as 
\begin{subequations}
	\begin{equation}
		\label{eq:ne+mpa}
		\boldsymbol{e}_{\square}^\prime\leftarrow\operatorname{AGG}_r^e\left(\varphi_m^{e}(\boldsymbol{v}_r,\boldsymbol{e}_\square,\boldsymbol{e}_{\square,r})\right),
	\end{equation}
	and
	\begin{equation}
		\label{eq:ne+mpb}
		\boldsymbol{v}_i\leftarrow\operatorname{AGG}_{\square}^v\left(\varphi_m^{v}(\boldsymbol{v}_i,\boldsymbol{e}_{\square_i}^\prime,\boldsymbol{e}_{\square_i,i})\right)
	\end{equation}
\end{subequations}
respectively, in which $\varphi_m^{e}$ and $\varphi_m^{v}$ are the element and node feature update function respectively, $\operatorname{AGG}$ denotes the mean aggregation function, $r=i,j,k,l$ denotes the index of the four nodes connected by the element, $\square_i$ denotes any element that connects node $i$ with other nodes, and $\boldsymbol{e}_{\square}^\prime$ denotes the element feature vector after the element update stage (Eq. \ref{eq:ne+mpa}).

As mentioned in Sec. \ref{sec:discretization}, we only need network output on the element-node edges, and therefore only an element-node decoder is needed
\begin{equation}
	\boldsymbol{e}_{\square,i}\leftarrow h^{ev}(\boldsymbol{v}_i,\boldsymbol{e}_{\square},\boldsymbol{e}_{\square,i}).
\end{equation}
that generates output element-node feature vector $\boldsymbol{e}_{\square,i}$ used in the time stepping of the flow field predictions, discussed later in Sec. \ref{sec:timestep}.

\paragraph*{Multi-layer perceptron}
\label{sec:mlp}
All the functions embedded within the hypergraph neural network, including the encoding functions $g^v$, $g^e$, and $g^{ev}$, element update functions $\varphi^e_m$, node update functions $\varphi^v_m$, and the decoding function $h^{ev}$, are chosen as multi-layer perceptrons with two hidden layers in this work, which can be explicitly written as a function $f$ to the input vector $z$
\begin{equation}
	f(\boldsymbol{z}) = \boldsymbol{W}_3\sigma_2(\boldsymbol{W}_2\sigma_1(\boldsymbol{W}_1\boldsymbol{z}+\boldsymbol{b}_1)+\boldsymbol{b}_2)+\boldsymbol{b}_3
\end{equation}
with weight matrices $\boldsymbol{W}_i$ and bias vectors $\boldsymbol{b}_i$ trainable for $i=1,2,3$. The number of rows in the weight matrix $\boldsymbol{W}_i$ is defined as the layer width, and the nonlinear function $\sigma_i$ is usually called the activation function. 

\subsection{Mesh adaptation}
\label{sec:meshadapt}
In most cases, the rotating motion of the solid body will inevitably lead to mesh distortion or adaptation. For computational fluid dynamics (CFD) simulations of such a fluid system, the sliding mesh approach \cite{murphy1994cfd} is frequently employed, with which the domain is partitioned into two parts, one static and the other co-rotating with the rigid body. Interpolation is performed on the interface at every time step. When a graph neural network is used, however, such interpolation becomes difficult. Information embedded as high-dimensional latent feature vectors propagates through the interface between the rotating and the static sub-domain with every network layer, leading to significant interpolation overhead. In addition, these information has no explicit physical meaning, making it difficult to determine how the interpolation should be performed. It is therefore more preferable to completely avoid any interpolation. For such a purpose, we adopt an additional layer of mesh \cite{behr1999shear} to handle the interface between the rotating and static sub-domains. This additional layer of mesh is allowed to deform over time, with their connectivity adapted when the deformation goes over a certain threshold. We lay out the implementation details in the remainder of this subsection.

\begin{figure}[t]
	\centering
	\includegraphics[]{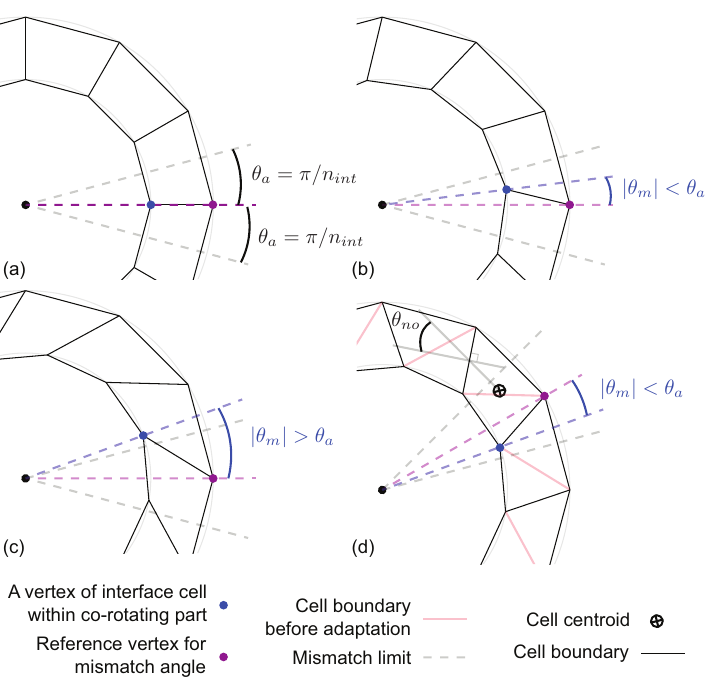}
	\caption{Schematic of the construction and adaptation of the interface cells. (a) Initial construction of the cells. (b) Rotation of the solid body leads to some distortion of the cells, as mismatch angle $\theta_m$ is smaller than the limit $\theta_a$ no mesh adaptation is performed. (c) Further solid body rotation cause more distortion of the cells, triggering the mesh adaptation, leading to the new interface cells in (d).}
	\label{fig:domainMesh}
\end{figure}

\paragraph*{Initialization}
For ease of illustration, we assume a 2D case here. Extension to 3D cases will be covered later in this subsection. Assume that the co-rotating sub-domain is a circle of radius $R$. Assume that we know the minimum mesh resolution needed to resolve the physics everywhere on this circle, and one of the appropriate mesh sizes is equal to a square with border length $2\pi R/n_{int}$, where $n_{int}$ is an integer and satisfies $n_{int}\geq4$. To approximately satisfy such mesh resolution at the interface, we partition the circle with radius $R$ by $n_{int}$ equi-distant vertices, construct a new circle that is concentric with the co-rotating sub-domain with radius $R+2\pi R/n_{int}$, and partition the new circle in the same way with another $n_{int}$ equi-distant vertices. Connecting these partitioning points on the two circles lead to $n_{int}$ quadrilateral cells between the two circles, like the ones shown in Fig. \ref{fig:domainMesh}a, which we denote as the interface cells. The other parts of the domain can be meshed normally with the two circles being treated as domain boundaries and the vertices on the two circles as the vertices on the domain boundaries.

\paragraph*{Mesh movement, connectivity adaptation, mesh quality}
Similar to that of the sliding mesh method, the vertices within the co-rotating sub-domain rotates along with the solid body, whilst the vertices within the static sub-domain does not, leading to distortion of the interface cells over the time steps, as shown in Fig. \ref{fig:domainMesh}b. For the purpose of maintaining mesh quality, the connectivity of the interface cells are adapted when such adaptation increases the mesh quality. For the specific interface cell construction adopted in this work, such connectivity adaptation happens every time the mismatch in angle $\theta_m$ within any of these cells is greater than $\theta_a=\pi/n_{int}$ rad, as is demonstrated in Fig. \ref{fig:domainMesh}c and \ref{fig:domainMesh}d. This construction bounds the non-orthogonality angle $\theta_{no}$, defined as the maximum angle between the normal vector of any cell boundary and the line connecting the cell centroid to the center of that cell boundary (marked out in Fig. \ref{fig:domainMesh}d) at the start of each time step to be
\begin{equation}
	\theta_{no}\leq\arcsin{\frac{\sin\theta_a\left[(2\theta_a+1)\cos\theta_a-2\theta_a(\theta_a+1)\right]}{\sqrt{(2\theta_a^2+2\theta_a+1)^2-(2\theta_a+1)^2\cos^2\theta_a}}}
\end{equation}
with limit $\lim\limits_{n_{int}\rightarrow\infty}{(\max(\theta_{no}))}=\arctan0.5\approx26.57^\circ$ when there are infinitely many interface cells.

\paragraph*{Extension to 3D}
The aforementioned 2D interface cell construction approach can be extended to 3D by extrusion, leading to semi-structured interface cells. Assume the extrusion direction is parallel to the $z$-axis. The cell size in $z$ direction does not have to be uniform, and the size the co-rotating sub-domain can also expand or shrink over the extrusion. The cells at each $z$ level can be treated as a separate 2D mesh and adapted level by level at every time step, using the same criterion as that of the 2D cases.

\section{Implementation of the surrogate model}
\label{sec:framework}
\begin{figure}
	\centering
	\includegraphics[]{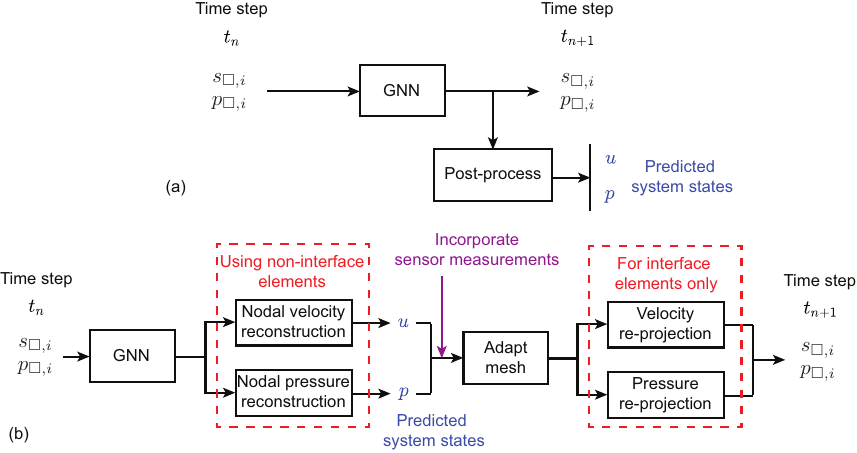}
	\caption{(a) Schematic of the data flow of the graph neural network for cases with time-invariant meshes with $\phi$-GNN in \cite{gao2024finite}. (b) Schematic of the data flow for the current framework.}
	\label{fig:dataflow}
\end{figure}

In this section, we assemble the $\phi$-GNN (Sec. \ref{sec:model}) with the mesh/graph adaptation method described in Sec. \ref{sec:meshadapt} to form a complete framework that can serve as an efficient surrogate model for fluid flow around rotating structures. We would start by transforming the mesh and system states into features (Sec. \ref{sec:featvec}), followed by a brief description on how the neural network predictions are generated over the time steps (Sec. \ref{sec:timestep}). After that, the nodal velocity and pressure values are reconstructed in Sec. \ref{sec:postproc} before the mesh movement/adaptation and re-projected to element-node features after the mesh movement/adaptation. A schematic of the entire data flow is provided in Fig. \ref{fig:dataflow}.

\subsection{Feature transformation, graph feature attachment}
\label{sec:featvec}
We consider the fluid system defined in Sec. \ref{sec:discretization}. We assume that flow velocity $\boldsymbol{u}=(u_x,u_y,u_z)$ and pressure $p$ is available at each vertex within the mesh, along with the coordinate of the vertex itself $(x,y,z)$. For 2D cases, we assume the $z$-component of all velocity vectors and vertex coordinates are all zero and can be neglected. We assume that the angular velocity of the solid structure $\omega_{t_n}$ and the accumulated rotation angle $\alpha_{t_n}$ are known at every single time step $t_n$. The mesh is assumed to be body-fitting, i.e., each system boundary is defined as a set of mesh cell boundaries, with boundary conditions ${\Gamma}$ known. The input features to the neural network are transformed from these information. 

\begin{figure}
	\centering
	\includegraphics[]{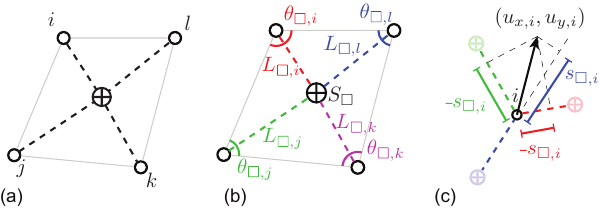}
	\caption{Schematic of geometry and flow feature transformations for 2D case: (a) a (quadrilateral) element $\square$ connecting four nodes, (b) the geometry features used, and (c) projection of flow velocity $(u_{x,i},u_{y,i})$ onto the direction of the local coordinates of the element-node edges. Figure modified from Fig. 4 in \cite{gao2024finite}.}
	\label{fig:geometry}
\end{figure}

\paragraph*{Geometry features}
We first transform the geometry features, in particular the vertex coordinates and the connectivity between the vertices, to graph features that are invariant to translation and rotation of the system. For an element denoted by a square symbol $\square$, one calculate the local coordinate of the nodes relative to the center of the cell
\begin{equation}
	\boldsymbol{x}_{\square,i}=(x_{\square,i},y_{\square,i},z_{\square,i})=(x_i-\overline{x_i},y_i-\overline{y_i},z_i-\overline{z_i}).
\end{equation}
with the overline $\overline{(\cdot)}$ denoting the mean over all nodes $i$ connected by the element $\square$. Using the length of the local coordinate vectors $L_{\square,i}=\sqrt{x_{\square,i}^2+y_{\square,i}^2+z_{\square,i}^2}$, the angles (2D) or solid angles (3D) at all corners of the element $\theta_{\square,i}$, and the area (2D) or volume (3D) of the element $S_\square$, we fully constrain the shape and size of the element. As these features do not constrain the location and rotation of the element, the desired translation and rotation invariance properties are satisfied, and therefore these information are suitable to be attached as features. A schematic of the geometry features is shown in Fig. \ref{fig:geometry}.

\paragraph*{Fluid flow features}
The fluid flow information available are the flow velocity $(u_x,u_y,u_z)$ and pressure $p$ at each node, as well as the set of boundary conditions ${\Gamma}$. With the geometry features being translation and rotation invariant, the scalar value of pressure can usually be used directly. The flow velocity, on the other hand, requires special transformations. Similar to the approach adopted in references \cite{lino2022multi,gao2024finite} for 2D cases, we project the flow velocity vector at each vertex onto the direction of the unit local coordinate vectors $\boldsymbol{x}_{\square,i}$ of the node, 
\begin{equation}
	\label{eq:proj}
	s_{\square,i} = \begin{bmatrix}
		u_{x,i}&u_{y,i}&u_{z_i}
	\end{bmatrix} \boldsymbol{x}_{\square,i} = \frac{1}{L_{\square,i}}\begin{bmatrix}
		u_{x,i}&u_{y,i}&u_{z_i}
	\end{bmatrix}\begin{bmatrix}
		x_{\square,i}\\
		y_{\square,i}\\
		z_{\square,i}
	\end{bmatrix},
\end{equation}
as illustrated in Fig. \ref{fig:geometry}c. The weights of these projections $s_{\square,i}$ are suitable to be used as features. Following the practice of Pfaff et al. \cite{pfaff2020learning}, we transform the boundary conditions into a one-hot vector $\boldsymbol{\gamma}_i$ for each node $i$.

\paragraph*{Graph feature attachment}
After the initial transformation discussed in earlier parts of this subsection, most of the geometric and flow features are already defined on element-node edges. We further gather the boundary condition and the pressure features from each of the nodes to all connected element-node edges,
\begin{subequations}
\begin{equation}
	\label{eq:scalarproj}
	\begin{aligned}
		p_{\square,i}&=p_i\\
		\boldsymbol{\gamma}_{\square,i}&=\boldsymbol{\gamma}_i,
	\end{aligned}
\end{equation}
and duplicate the element area/volume feature from the element to all connected element-node edges,
\begin{equation}
	S_{\square,i}=S_\square.
\end{equation}
\end{subequations}
The processed features are then concatenated into the element-node edge feature vectors,
\begin{equation}
	\label{eq:nefeatcyl}
	\boldsymbol{e}_{\square,i}=\begin{bmatrix}
		s_{\square,i}&
		p_{\square,i}&
		\ln L_{\square,i}&
		\ln S_{\square,i}&
		\cos \theta_{\square,i}&
		\boldsymbol{\gamma}_{\square,i}&
		\operatorname{par}_{\square,i}
	\end{bmatrix}^T.
\end{equation} 
where $\operatorname{par}_{\square,i}$ denotes any additional information of the system, which is the maximum angle of attack for the 2D experiment reported in Sec. \ref{sec:results2d}.

\subsection{Time stepping}
\label{sec:timestep}
As discussed in the earlier parts of this work, the graph neural network is used as a surrogate model $\widehat{\boldsymbol{G}}$ for the ground truth feature update function in Eq. \ref{eq:graphstep}. Specifically, the velocity and pressure features embedded on the element-node edge feature vectors are iteratively updated by the graph neural network
\begin{equation}
	\label{eq:timestep}
	\begin{aligned}
		\boldsymbol{e}_{\square,i}^{t_n+1}&=\boldsymbol{e}_{\square,i}^{t_n}+\begin{bmatrix}
			s_{\square,i}^{t_n+1}-s_{\square,i}^{t_n}\\
			p_{\square,i}^{t_n+1}-p_{\square,i}^{t_n}\\
			\boldsymbol{0}_{5\times1}
		\end{bmatrix}\\
		&\approx \boldsymbol{e}_{\square,i}^{t_n}+\begin{bmatrix}
			\widehat{\boldsymbol{G}}(\boldsymbol{e}_{\square,i}^{t_n})\\
			\boldsymbol{0}_{5\times1}
		\end{bmatrix}
	\end{aligned}
\end{equation}

\subsection{Reconstruction and re-projection of velocity and pressure}
\label{sec:postproc}
The movement of the rotating sub-domain changes the geometry of the interface cells at every time step, and changes the mesh topology every a few time steps. The velocity feature used as the network input, constructed with Eq. \ref{eq:proj}, implicitly assumes invariant cell shape, and therefore the velocity features projected on the element-node edges within the interface elements will become inaccurate after the mesh update and have to be discarded after every time step. As the velocity features embedded on the element-node edges connected to the interface elements are still needed as the network input in the subsequent time step, we will have to re-calculate the correct values for these features. A similar treatment is also needed for the pressure features as the treatment in Eq. \ref{eq:scalarproj} also assumes invariant graph connectivity. The re-calculations are achieved in a two-stage manner.

\begin{figure}[t]
	\centering
	\includegraphics[]{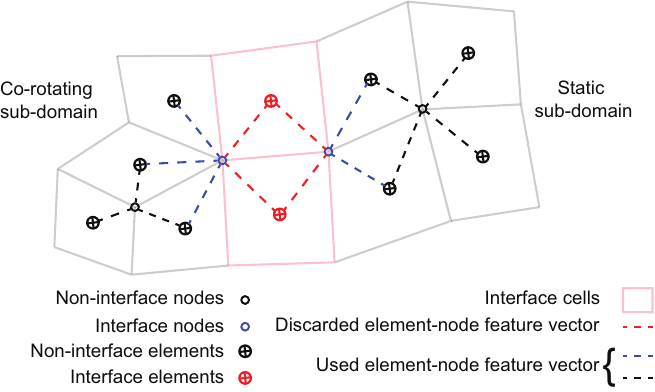}
	\caption{Schematic indicating the used and discarded element-node edge features in the nodal velocity and pressure reconstruction process.}
	\label{fig:velRecon}
\end{figure}

\paragraph*{Velocity and pressure reconstruction}
In the first stage, the nodal velocity vectors and the nodal pressure values are reconstructed with all the element-node vectors belonging to the interface elements discarded, as is shown in Fig. \ref{fig:velRecon}. 
The velocity at each node is obtained by solving a Moore-Penrose pseudo-inverse \cite{lino2022multi,gao2024finite}, using the prediction velocity features as inputs
\begin{equation}
	\label{eq:invgeoproj}
	\begin{aligned}
		\hat{\boldsymbol{u}}_i^{t_n}=&\begin{bmatrix}
			\hat{u}_{x,i}^{t_n}\\
			\hat{u}_{y,i}^{t_n}\\
			\hat{u}_{z,i}^{t_n}\\
		\end{bmatrix}\\
		=&\left((\boldsymbol{1}_{3\times1}(\hat{\boldsymbol{s}}_{\square_i,i}^{t_n})^T)\odot((\boldsymbol{X}_{\square_i,i}\boldsymbol{X}_{\square_i,i}^T)^{-1}\boldsymbol{X}_{\square_i,i})\right)\boldsymbol{1}_{n_{\square_i,i}\times1},
	\end{aligned}
\end{equation}
where the symbol $\odot$ denotes the Hadamard product, $n_{\square_i,i}$ is the number of elements connecting node $i$ with other nodes, $\hat{\boldsymbol{d}}_{\square_i,i}$ is a $n_{\square_i,i}\times1$ vector containing the predicted projection weights $\hat{d}_{\square_i,i}$ attached to all the element-node edges connected to node $i$, and $\boldsymbol{X}_{\square_i,i}$
is a $3\times n_{\square_i,i}$ matrix containing the unit local coordinate vectors $\boldsymbol{x}_{\square_i,i}$ of node $i$ within all the elements $\square_i$ that connect node $i$ with other nodes. 
For pressure, the value at each node is obtained by evaluating the mean across all element-node edges connecting the node
\begin{subequations}
	\label{eq:scatterp}
	\begin{equation}
		\hat{p}_i^n=\operatorname{MEAN}_\square(\hat{p}_{\square_i,i}^n),
	\end{equation}
\end{subequations}
in which $\operatorname{MEAN}_\square$ is the mean aggregation function. The reconstructed nodal velocity and pressure values also serve as the network prediction output, so no additional post-processing steps are needed, as shown in Fig. \ref{fig:dataflow}b.

\begin{remark}
	One might notice that the pseudo-inverse in Eq. \ref{eq:invgeoproj} is different at every time step due to the rotation of the co-rotating sub-domain. Such re-calculation turns out to be computationally too expensive, as a linear system needs to be solved for every single node within the graph, and the size of such linear system varies for different nodes. In implementation, we avoid the calculation of pseudo-inverses at every time step by taking advantage of the rotation equivariance of the velocity features -- We take track of the accumulated rotation angle $\alpha_{t_n}$ of the solid body, reconstruct the velocity field using the pseudo-inverse values calculated from the initial reference mesh, and then rotate all the reconstructed velocity vectors within the co-rotating sub-domain by $\alpha_{t_n}$.
\end{remark}

\paragraph*{Velocity and pressure re-projection}
The second stage of the interface element-node feature re-calculation takes place after the mesh adaptation. The geometry features of the interface cells are re-calculated from the nodal coordinates, along with the local coordinate vectors. After that, the reconstructed nodal velocity vectors are projected to the new local coordinate vectors using Eq. \ref{eq:proj} to obtain the new velocity feature vector for the interface elements. The pressure features are obtained through Eq. \ref{eq:scalarproj} using the new interface element connectivity.

\section{Experimental setup}
\label{sec:setup}
With the framework constructed in Sec. \ref{sec:framework}, we proceed to apply it to fluid systems around rotating structures. For this purpose, we consider two test cases. The setup of these cases will be discussed in the subsequent Sec. \ref{sec:case}, including the CFD setup and also how the training, cross-validation and test data sets are obtained.

\subsection{Setup of the test cases}
\label{sec:case}
We test our proposed framework on two test cases, one in 2D and the other in 3D. The 2D test case features an oscillating NACA0012 airfoil in uniform flow. The 3D test case involves the flow around a rotating right cube. In the remainder of this subsection, we will describe how the CFD simulations are performed and validated, how the flow data are sampled, and how these flow data are pre-processed to obtain the training and test data sets for the two test cases.

\paragraph*{Solver setup}
We use open-source package OpenFOAM \cite{weller1998tensorial} to generate ground-truth CFD data for both test cases. The Navier-Stokes equations are discretized using the finite-volume method. The time derivatives are discretized by a second-order implicit scheme, and the divergence, gradients, and diffusion terms are discretized via a second-order Gaussian integration scheme. The discretized equations are solved by the Pressure-Implicit with Splitting of Operators (PISO) algorithm \cite{issa1986solution} with one non-orthogonal corrector. The tolerance for both velocity and pressure residuals are set to be $1\times10^{-8}$. We validate the numerical scheme by simulating the flow around a static cube at $Re=200$. The average drag coefficient is calculated to be $c_d=0.935$ in that validation case, which is between the values obtained from empirical formulas by H{\"{o}}lzer et al. \cite{holzer2008new} ($c_d=0.854$) and Haider and Levenspiel \cite{haider1989drag} ($c_d=1.122$). We therefore believe that the simulation results are reasonable. 

\begin{figure}[t]
	\centering
	\includegraphics[]{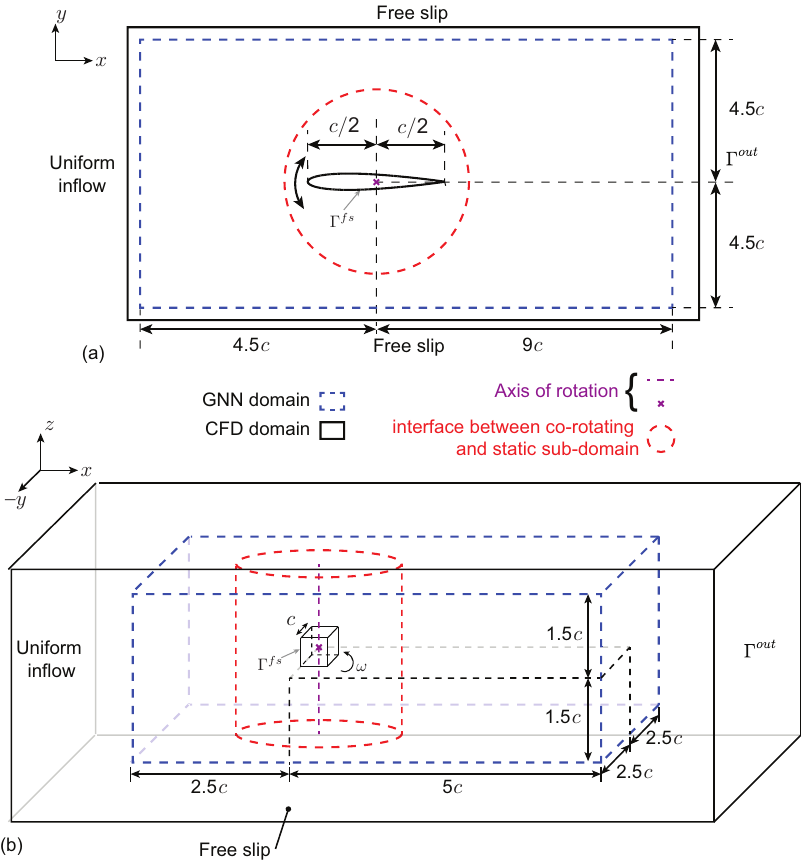}
	\caption{Schematic of CFD and GNN computational domains for (a) the 2D flow around oscillating airfoil test case, and (b) the 3D flow around rotating cube test case.}
	\label{fig:domains}
\end{figure}

\begin{figure}
	\centering
	\includegraphics[]{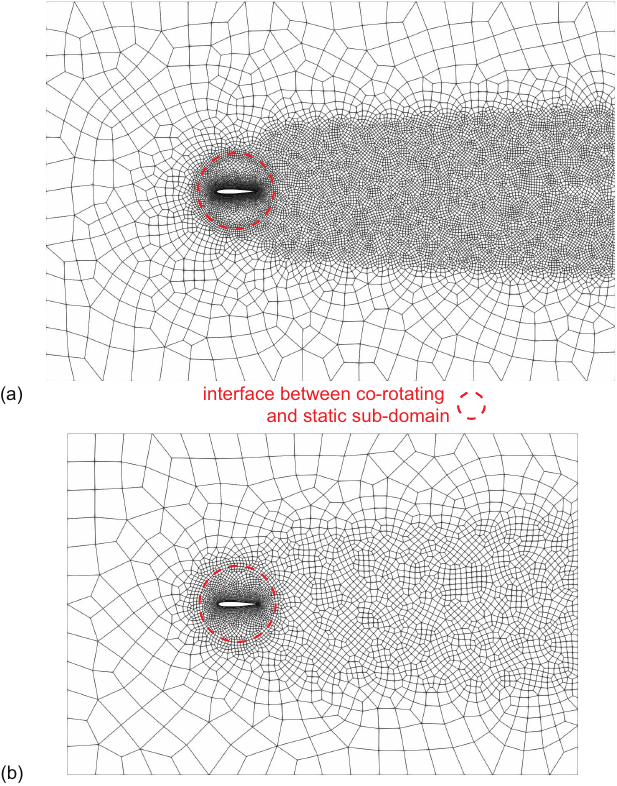}
	\caption{(a) Mesh used for the CFD simulation, and (b) mesh used by the GNN surrogate model, for the 2D flow around oscillating airfoil test case. The red dashed circle marks out the interface between the co-rotating and static sub-domains.}
	\label{fig:2dmesh}
\end{figure}

\paragraph*{2D case: flow around oscillating airfoil}
A schematic of the simulation domain for the 2D test case is shown in Fig. \ref{fig:domains}a. A NACA0012 airfoil with chord length $c$ oscillates around the center of its chord line, which is placed at the origin. The oscillation follows a sinusoidal pattern, with the angle of attack being
\begin{equation}
	\operatorname{AOA} = \operatorname{AOA}_{\max}\sin(2\pi t^*)
\end{equation}
with $\operatorname{AOA}_{\max}$ varying in a large range. The system domain $\Omega$ is rectangular, with its boundaries shown in Fig. \ref{fig:2dmesh}a. The inlet, the top, and the bottom boundaries are $5c$ away from the origin, whilst the outlet is $10c$ away from the origin. No-slip boundary condition is applied to the surface $\Gamma^{fs}$ of the airfoil. Free-slip boundary condition is applied to the top and bottom domain boundaries. The inlet features a uniform flow $\boldsymbol{U}_\infty=(U_\infty,0,0)$, whilst a Dirichlet total pressure boundary condition is applied to the outlet $\Gamma^{out}$. The radius of the co-rotating sub-domain is $c$. The Reynolds number is fixed at $\operatorname{Re}_c=2000$.

With the mesh shown in Fig. \ref{fig:2dmesh}a, we generate a series of simulations with maximum angle of attack $\operatorname{AOA}_{\max}\in[0,30]$ that are used to train, cross-validate, and test the model. In particular, we use the simulations with $\operatorname{AOA}_{\max}=5^\circ,7^\circ,\ldots,23^\circ,25^\circ$ to construct the training and cross-validation sets, and the simulations with $\operatorname{AOA}_{\max}=0^\circ,1^\circ,2^\circ,3^\circ,4^\circ,6^\circ,8^\circ,\ldots,24^\circ,26^\circ,27^\circ,28^\circ,29^\circ,30^\circ$ to construct the test data sets. All these simulations are performed with maximum time step size $\delta t^*=\delta tU_\infty/c=0.0008$, which corresponds to Courant number limit about $\operatorname{Co}\approx0.95$ for cases with large oscillation amplitudes, and about $\operatorname{Co}\approx0.15$ for cases with small oscillation amplitudes. We sample the train and test data sets at each Reynolds number after the simulation reaches the statistically stationary state, with the non-dimensionalized sampling time step $\Delta t^*=\Delta tU_\infty/c=0.04$, equivalent to $50$ CFD time steps. For each train/cross-validate simulation, we sample 2049 continuous time steps. For each test simulation, we sample 1001 continuous time steps. The flow data samples are loaded via the FluidFOAM package \cite{augier2019fluiddyn} and subsequently interpolated onto the mesh drawn in Fig. \ref{fig:2dmesh}b using a linear 2D interpolation available in the SciPy package \cite{virtanen2020scipy}. The interpolated flow data are then processed into training, cross-validation, and test data sets, with details presented later in this subsection.

\begin{figure}
	\centering
	\includegraphics[]{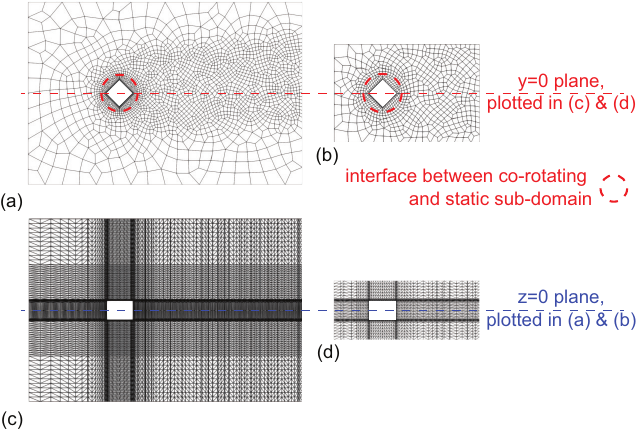}
	\caption{(a) \& (c) Mesh used for the CFD simulation, and (b) \& (d) mesh used by the GNN surrogate model, for the 3D flow around rotating cube test case. The red dashed circle marks out the interface between the co-rotating and static sub-domains. The actual meshes only contain hexahedral cells recombined from the ones shown in (c) \& (d).}
	\label{fig:3dmesh}
\end{figure}

\paragraph*{3D case: flow around a rotating cube}
It should be noted that the proposed framework can also be applied to 3D scenarios. We demonstrate such capability using a 3D case of the flow around a rotating cube. A schematic of the domain $\Omega$ for the 3D test case is shown in Fig. \ref{fig:domains}b. A cube is placed with its center at the origin, and its six faces parallel to $x\pm y=0$ and $\pm z$ planes respectively at $t^*=0$. The cube rotates around $+z$-axis with non-dimensionalized angular velocity $\omega^*=\omega c/U_\infty=0.25$, where $U_\infty$ is the inlet flow velocity, $\omega$ is the angular velocity, and $c$ equals to the border length of the cube as well as the radius of the co-rotating sub-domain. The simulation domain $\Omega$ is defined as a cuboid that contains the cube. Similar to the 2D case of flow around oscillating airfoil, we assign a uniform flow $\boldsymbol{U}_\infty=(U_\infty,0,0)$ to the inlet, a Dirichlet total pressure boundary condition to the outlet, and free-slip boundary condition to the other four faces of the cuboid. No-slip boundary condition is assigned to the surface $\Gamma^{fs}$ of the cube. With the mesh plotted in Fig. \ref{fig:3dmesh}a and \ref{fig:3dmesh}c, we generate simulation data at Reynolds number $Re=100$ with a non-dimensionalized time step $\delta t^*=\delta t U_\infty/c=5\times10^{-3}$, which corresponds to Courant number $\operatorname{Co}\approx0.56$. After the simulation reaches stationary state, we sample 2049 continuous time steps with non-dimensionalized sampling time step $\Delta t^*=\Delta tU_\infty/c=0.04$ (i.e., every 8 CFD time steps) that are then processed into training and cross-validation data set, and another (without overlapping with the training set) 1001 continuous time steps at the same interval $\Delta t^*=0.04$ for test data set. The sampled and loaded flow data are interpolated onto the mesh drawn in Fig. \ref{fig:3dmesh}b and \ref{fig:3dmesh}d using a radial basis function interpolation. Note that both meshes are semi-structured (unstructured in $x-y$ plane and structured in $z$-axis direction). The interpolated flow data are then processed into training, cross-validation and test data sets.

\paragraph*{Construction of training, cross-validation, and test data sets}
With the flow data sampled, we proceed to construct the data sets. We construct the training samples following Eq. \ref{eq:timestep}, meaning that the 2049 continuous time steps sampled from each of the training simulations can be separated into 2048 training samples, with each sample containing input-output pair 
\begin{equation}
	\label{eq:sample}
	\begin{aligned}
		\text{Input:}&\quad \boldsymbol{e}_{\square,i}^{t_n}\\
		\text{Output:}&\quad \begin{bmatrix}
			s_{\square,i}^{t_n+1}-s_{\square,i}^{t_n}\\
			p_{\square,i}^{t_n+1}-p_{\square,i}^{t_n}
		\end{bmatrix}\\
	\end{aligned}
\end{equation}
for time step $n=1,2,\ldots,2048$. The training samples are then concatenated to form the training set. The input element-node edge features and output element-node edge features are then normalized via min-max normalization with minimum $-1$ and maximum $1$, respectively, for the entire training set. For cross-validation and test data sets, we construct the inputs to the neural network using the first time step of the sampled data in the same way as we construct the training input, and then normalize such inputs using the normalization statistics (i.e., the scaling and shift factors) calculated from the training data sets.

\subsection{Network setup and training}
\label{sec:setupTr}
The networks are implemented via PyTorch \cite{paszke2019pytorch}. A $\phi$-GNN with $15$ layers is used to model the 2D rotating mixer case, and another $\phi$-GNN with $10$ layers is used for the 3D rotating cube case. For both networks, the encoding, decoding, as well as the node and element update functions are all selected as multi-layer perceptrons with two hidden layers of width $192$, activated by sinusoidal activation function \cite{sitzmann2020implicit}. All aggregation functions are chosen as the mean function. 

\paragraph*{Hardware settings} All training and tests are performed with a single Nvidia RTX 3090 GPU with CPU being AMD Ryzen 9 5900 @ 3.00 GHz$\times$12 cores. We cap the CPU frequency at 2.91 GHz during training of the neural network to avoid performance fluctuations. The GPU is downclocked by 593 MHz and limited to 75\% power during training for the same reason. The test runs are performed without these limitations. We report the training and inference speed along with the GPU memory overhead in Table \ref{tab:speed}.

\begin{table}[]
	\caption{Various statistics for the test cases. The peak GPU memory overhead occurs during training, memory overhead during inference is very low. Inference time for GNN includes all overhead needed for running the network itself, mesh adaptation, velocity reconstruction \& re-projection, and saving prediction output. The 2D CFD is run with a single CPU core, whilst the 3D CFD case is run in parallel with 12 CPU cores. The GNN framework for both 2D and 3D cases runs on a single RTX 3090 GPU.}
	\label{tab:speed}
	\centering
	\begin{tabular}{cccccc}
		\toprule
		Case & Type & \begin{tabular}[c]{@{}c@{}}Mesh size\\ (number of \\ cells/elements)\end{tabular} & \begin{tabular}[c]{@{}c@{}}Train speed\\ (min/epoch)\end{tabular} & \begin{tabular}[c]{@{}c@{}}Peak GPU \\ Memory\\ (GB)\end{tabular} & \begin{tabular}[c]{@{}c@{}}Inference speed\\ (ms/GNN step)\end{tabular} \\ \midrule
		\multirow{2}{*}{2D} & CFD & \begin{tabular}[c]{@{}c@{}}15476\\ (pure quad)\end{tabular}  & - & - & 7847 (156.9$\times$50) \\
		& GNN & \begin{tabular}[c]{@{}c@{}}4630\\ (pure quad)\end{tabular}  & $\approx16$ & 7.4 & 38.1 \\ \midrule
		\multirow{2}{*}{3D} & CFD & \begin{tabular}[c]{@{}c@{}}464355\\ (pure hex)\end{tabular} & - & - & 8871 (1108.9$\times$8) \\
		& GNN & {\begin{tabular}[c]{@{}c@{}}32776\\ (pure hex)\end{tabular}} & $\approx18$ & 15.9 & 165 \\ \bottomrule
	\end{tabular}
\end{table}

\paragraph*{Training scheme \& loss} An Adam \cite{kingma2014adam} optimizer with $\beta_1=0.9$ and $\beta_2=0.999$ is used for the training. The network is trained for 100 epochs for the 2D case with batch size 4, and 200 epochs for the 3D case with batch size 1 due to a smaller data set size. For both cases, the learning rate is adjusted at the start of every epoch following a merged cosine-exponential scheme \cite{gao2024finite} with maximum learning rate $1\times10^{-4}$ and minimum learning rate $1\times10^{-6}$. An adaptive smooth $L_1$ loss \cite{gao2024finite} is used, which calculates the loss for ground truth $\psi$ and its neural network prediction $\hat{\psi}$ as 
\begin{equation}
	\begin{aligned}
		L_i(\psi_i,\hat{\psi}_i) = 
		\begin{cases}
			(\psi_i-\hat{\psi}_i)^2/2\beta,&\text{if }\lvert\psi_i-\hat{\psi}_i\rvert<\beta\\
			\lvert\psi_i-\hat{\psi}_i\rvert-\beta/2,  &\text{if }\lvert\psi_i-\hat{\psi}_i\rvert\geq\beta\\
		\end{cases}
	\end{aligned}
\end{equation}
where the subscript $(\cdot)_i$ denotes entry-by-entry calculation, and $\beta$ controls the transition point between the $L_2$ loss region and the $L_1$ loss region, updated after each training iteration with
\begin{equation}
	\label{eq:betaUpdate}
	\beta^2\leftarrow (1-\frac{1}{N_{b}})\beta^2+\frac{1}{N_{b}}\min\{\beta^2,\operatorname{MSE}(\psi_{batch},\hat{\psi}_{batch})\}
\end{equation}
in which $N_b$ is the total number of training iterations within each epoch, and $\operatorname{MSE}$ denotes the mean-squared loss. Since the $\beta$ value is updated on-the-fly, its initial value only has a minor impact on the final results as long as such value is not set too close to zero (as the loss would reduce to an $L_1$ loss in that case). In this study, we manually set this initial value as $\beta=5\times10^{-3}$, close to the $L_2$ loss at the start of the training for the two cases.

\paragraph*{Training noise}
Assuming normally-distributed single-step prediction error at each time step, we adopt the training noise scheme from \cite{gao2024finite}, with which we add a Gaussian noise of $(\boldsymbol{\kappa}\odot\widehat{\boldsymbol{\epsilon}}_{\square,i}^n$,$-\omega\widehat{\boldsymbol{\epsilon}}_{\square,i}^{n})$ to each input-output pair (i.e., each training sample), where $\boldsymbol{\kappa}$ is the vector containing the scaling factors used during the min-max normalization of training output targets. The over-correction factor is set to $\omega=1.2$ for the 2D case and $\omega=1.0$ for the 3D case. Each entry of $\widehat{\boldsymbol{\epsilon}}_{\square,i}^{n}$ is randomly and individually sampled from a normal distribution, i.e., 
\begin{equation}
	\widehat{\boldsymbol{\epsilon}}_{\square,i}^{n}\sim\mathcal{N}(0,\operatorname{Var}(\widehat{\boldsymbol{\epsilon}}_{\square,i}^n))
\end{equation}
The variance of the noise adopted in this work are $\operatorname{Var}(\widehat{\boldsymbol{\epsilon}}_{\square,i}^n)=1\times10^{-2}$ for both cases, \emph{without any tuning}. We will show later in Sec. \ref{sec:results2d} that such lack of training noise optimization can lead to significant prediction error accumulation over the prediction rollout, but such error can be limited by the incorporation of the measurements from sparse pressure sensors, and thus nullifies the need for training noise tuning.

\subsection{Mixed precision training}
\label{sec:fp16}
To reduce memory overhead and improve inference speed, models are trained through mixed precision training, with which almost all calculations except the calculation of loss are performed with the floating-point number of half precision (FP16) rather than the typical floating-point number of single precision (FP32). Since half-precision floating-point numbers offer significantly lower precision compared with single precision ones, to prevent gradient underflow, we followed the standard practice \cite{micikevicius2018mixed} to scale up the loss by a multiplier $\lambda$ during the gradient calculation process, and the calculated gradients are scaled back afterwards. In this work, this multiplier starts with $\lambda=2^{11}=2048$, and doubles its value when $\lambda\beta<125$ is satisfied at the start of any epoch, where $\beta$ is the parameter updated on-the-fly in Eq. \ref{eq:betaUpdate}. The constant 125 is obtained from preliminary tests to ensure that the loss and the gradient do not overflow (i.e., the absolute value of their magnitude stays within 65504, the upper limit of half-precision floating-point numbers) for $\phi$-GNNs. Since the smallest positive number that can be expressed with half-precision floating-point number is around $6\times10^{-8}$, we also adjust the $\epsilon$ value in the denominators within the optimizer to $10^{-7}$ rather than the default $10^{-8}$ for the sake of numerical stability.

\subsection{Evaluation metrics}
\label{sec:metric}
We test the trained networks on the test data sets to evaluate their performance. Similar to existing works such as \cite{lino2022multi,gao2024finite}, we use both qualitative and quantitative metrics to facilitate evaluation. 
For the qualitative measure, we compare the velocity magnitude fields obtained from the predictions versus those from CFD simulations (i.e., the ground truth data). For the 2D test case, we will plot the whole domain. For the 3D test case, several 2D slices of the domain will be extracted for comparison.

A few quantitative measures are employed to facilitate comparison between the CFD results and the neural network predictions. To compare the similarity between predicted and ground truth fields, we adopt the coefficient of determination $R^2$, defined as 
\begin{equation}
	R^2=1-\frac{\|\boldsymbol{q}-\hat{\boldsymbol{q}}\|_2^2}{\|\boldsymbol{q}-\bar{\boldsymbol{q}}\|_2^2},
\end{equation}
for any system state $\boldsymbol{q}$ and its prediction $\hat{\boldsymbol{q}}$, where the overline $\overline{(\cdot)}$ denotes the mean operation. A coefficient of determination close to 1 indicates close matching between the predicted and ground-truth system states. We additionally calculate the forces on the solid body from the predicted system states and compare those with the forces calculated from the ground-truth system states. This calculation is achieved by integrating the Cauchy stress tensor $\boldsymbol{\sigma}=-p\boldsymbol{I}+\mu(\nabla\boldsymbol{u}+(\nabla\boldsymbol{u})^T)$ on the surface of the solid body $\Gamma_{fs}$ using the first layer of elements from that surface, where $\mu$ is the viscosity of the fluid. For the 2D test case, we calculate the drag (in the direction of $+x$-axis) and the lift (in the direction of the $+y$-axis) via
\begin{subequations}
	\label{eq:liftdrag}
	\begin{equation}
		C_l=\frac{1}{\frac{1}{2}\rho^f(U_c)^2A}\int_{\boldsymbol{\Gamma}^{fs}}(\boldsymbol{\sigma}^f\cdot \boldsymbol{n})\cdot \boldsymbol{n}_y d\boldsymbol{\Gamma},
	\end{equation}
	\begin{equation}
		C_d=\frac{1}{\frac{1}{2}\rho^f(U_c)^2A}\int_{\boldsymbol{\Gamma}^{fs}}(\boldsymbol{\sigma}^f\cdot \boldsymbol{n})\cdot \boldsymbol{n}_x d\boldsymbol{\Gamma}.
	\end{equation}
\end{subequations}
where $A=c$ for the 2D case and $A=c^2$ for the 3D case.

\begin{figure}[t]
	\centering
	\includegraphics[]{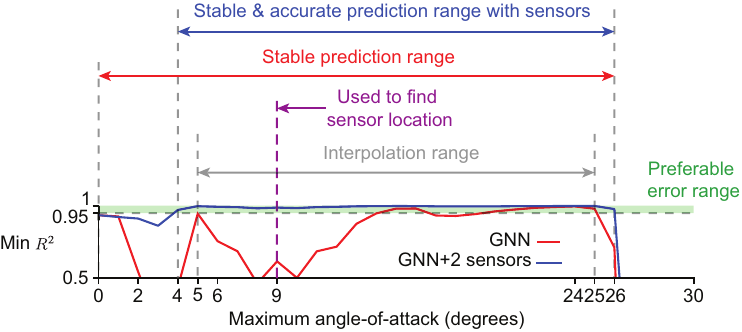}
	\caption{Overview of results for the 2D flow around oscillating airfoil test case. The minimum coefficient of determination $R^2$ over the prediction rollout from step 200 to 1000 are plotted for all the cross-validation and test sets over maximum angle-of-attack $\operatorname{AOA}_{\max}\in[0^\circ,30^\circ]$. We also mark out the range where the framework is able to generate stable predictions and the range where the integration of sensors enforces accurate prediction results (i.e., $R^2>0.95$ over the prediction rollout). }
	\label{fig:aoaregime}
\end{figure}

\begin{figure}[!t]
	\centering
	\includegraphics[]{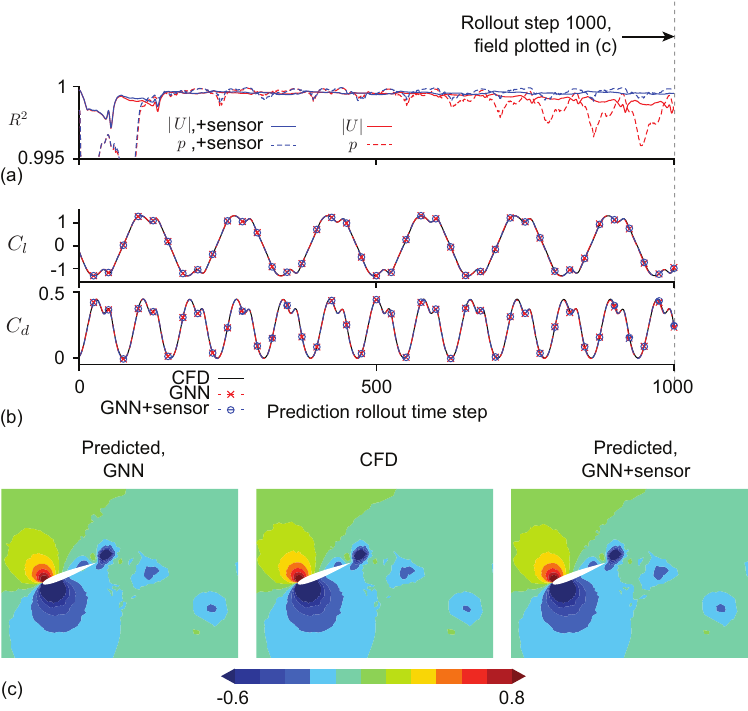}
	\caption{Results for the 2D flow around oscillating airfoil test case at $\operatorname{AOA}_{\max}=24^\circ$. (a) Coefficient of determination $R^2$ of the predicted velocity magnitude and pressure fields over the entire prediction rollout. (b) Lift and drag coefficients calculated from the predicted system states versus the CFD values. (c) Predicted non-dimensionalized pressure field $p^*$ near the airfoil at prediction rollout time step 1000, versus the CFD results.}
	\label{fig:results2d24}
\end{figure}

\section{Results and discussion}
\label{sec:results}
\subsection{Experiment -- 2D flow around oscillating airfoil}
\label{sec:results2d}
The network trained following the setup discussed in Sec. \ref{sec:setup} is tested on a series of test flow data sets obtained from simulations with $\operatorname{AOA}_{\max}=0^\circ,1^\circ,2^\circ,3^\circ,4^\circ,6^\circ,8^\circ,\ldots,24^\circ,26^\circ,27^\circ,28^\circ,29^\circ,30^\circ$, which covers the entire training $\operatorname{AOA}_{\max}$ range and also some extrapolation regime around it, as is shown in Fig. \ref{fig:aoaregime}. At each of the testing $\operatorname{AOA}_{\max}$, we start from the first time step within the test data set, and generate roll-out predictions of the system states (velocity and pressure) over the following 1000 time steps. We present these results and discuss the behavior of the network in the remainder of this subsection.

\begin{figure}[t]
	\centering
	\includegraphics[]{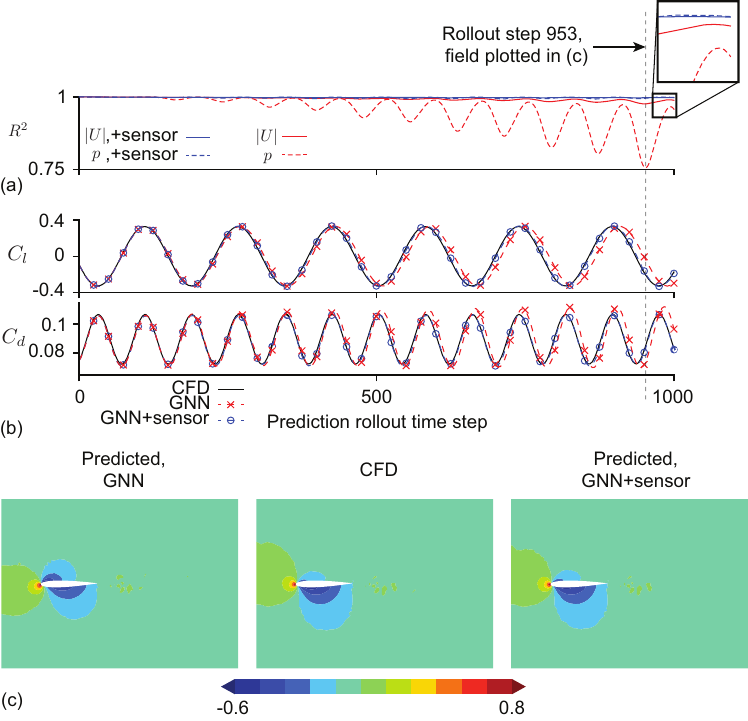}
	\caption{Results for the 2D flow around oscillating airfoil test case at $\operatorname{AOA}_{\max}=6^\circ$. (a) Coefficient of determination $R^2$ of the predicted velocity magnitude and pressure fields over the entire prediction rollout. (b) Lift and drag coefficients calculated from the predicted system states versus the CFD values. (c) Predicted non-dimensionalized pressure field $p^*$ near the airfoil at prediction rollout time step 953, versus the CFD results.}
	\label{fig:results2d6}
\end{figure}

\begin{figure}[t]
	\centering
	\includegraphics[]{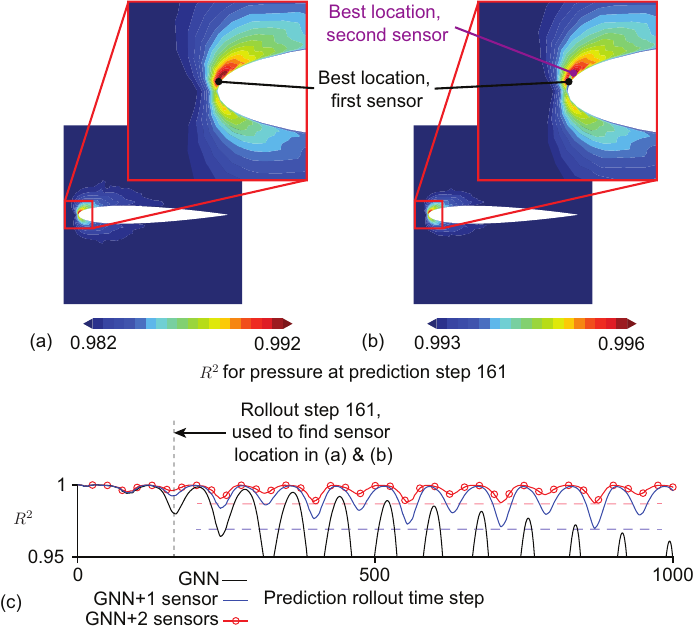}
	\caption{Finding the best sensor locations for the 2D flow around oscillating airfoil test case. (a) Coefficient of determination $R^2$ for pressure at prediction step 161 for all possible first sensor locations, with the best sensor location marked by a black dot. (b) Coefficient of determination $R^2$ for pressure at prediction step 161 for all possible second sensor locations, with the best sensor location marked by a magenta dot. (c) The coefficient of determination $R^2$ for pressure over the entire prediction roll-out for the training/cross-validation data set at $\operatorname{AOA}_{\max}=9^\circ$, with and without the use of sensors.}
	\label{fig:sensoraf}
\end{figure}

\paragraph*{Prediction within the interpolation range, accumulated phase and amplitude error}
The network generates stable prediction results throughout the interpolation range of $\operatorname{AOA}_{\max}\in[5^\circ,25^\circ]$, and also able to generate stable prediction results for a moderate extrapolation range of $\operatorname{AOA}_{\max}\in[0^\circ,26^\circ]$. The predictions are mostly accurate for higher values of $\operatorname{AOA}_{\max}$. We report one of such cases ($\operatorname{AOA}_{\max}=24^\circ$) in Fig. \ref{fig:results2d24}. However, for cases with lower $\operatorname{AOA}_{\max}$ values, the network predictions are not that accurate, with phase and amplitude errors accumulating over the prediction rollout. We plot the results of one of such cases ($\operatorname{AOA}_{\max}=6^\circ$) in Fig. \ref{fig:results2d6}. One could conclude from the $R^2$ as well as the lift and drag coefficient curves that prediction errors accumulate over the rollout, and reach non-negligible levels after a few hundred time steps. The accumulation of errors over the prediction rollout can partially be attributed to single-step supervision (cf. Eq. \ref{eq:sample}). Using multiple-step supervision will likely reduce the rate of error accumulation over prediction rollout \cite{list2024differentiability}, but at the same time lead to much slower neural network training and complicate the construction of artificial training noise \cite{sanchez2020learning} necessary for the stability of the predictions. The accumulated prediction error can also be partially attributed to the inherent bias in the training data -- the fluctuation in the velocity and pressure fields becomes more significant for larger $\operatorname{AOA}_{\max}$ values, causing the network to bias towards learning those dynamics during the training, leading to more accurate predictions for larger $\operatorname{AOA}_{\max}$ values and more error for smaller $\operatorname{AOA}_{\max}$ values. Such bias also makes it difficult to reduce the accumulated error by tuning the amount of added training noise. We therefore choose to neither use multiple-step supervision nor tune the amount of training noise. Instead, we notice that one could enforce accurate predictions by incorporating the measurements from sparse sensors, i.e., forming a digital twin model. In this work, we report these observations and the approach to find optimal sensor locations.

\begin{remark}
	One might notice from the $R^2$ curve in Fig. \ref{fig:results2d24}a that the prediction error was initially larger but the network self-corrects its prediction error after about the first 100 time steps. We have reported similar observations of self-correction in our previous work \cite{gao2024predicting}. The investigation of the exact reason behind this behavior is out of the scope of this research and requires a dedicated effort as is preferable as part of our future work.
\end{remark}

\paragraph*{Bounding accumulated prediction error with sparse sensors}
Formally, by integrating the measurements from sparse sensors, we seek to improve the accuracy of the predictions, quantified by the coefficient of determination $R^2$, by the most amount possible, using the least possible number of sensors. The integration of sensor measurements is achieved by replacing the predicted system states at the sensor locations with the measured ground truth values. To avoid interpolations, we restrict the possible sensor locations to all the nodes within the hypergraph. The best locations of the sensors are found iteratively. In each iteration, we fix the best sensor locations from previous iterations, and loop through all the possible new sensor locations (i.e., all the other nodes), record the $R^2$ value with each possible sensor placement at the end of a short prediction rollout, choose the node with the best $R^2$ value as the best sensor location, and start the next iteration. 
For this 2D flow around oscillating airfoil case, we use the train/cross-validation data set at $\operatorname{AOA}_{\max}=9^\circ$ to determine the best sensor locations. This data set is chosen because it features the worst accumulated prediction error within the training/cross-validation data sets. Note that we use a training data set because it is not proper to use test data sets other than testing the framework. The first two iterations give us two sensor locations rather close to the leading edge of the airfoil, plotted in Fig. \ref{fig:sensoraf}a and \ref{fig:sensoraf}b. Figure \ref{fig:sensoraf}c shows the $R^2$ values over the entire prediction roll-out. It can be clearly observed that the use of a single sensor already significantly slows down the accumulation of prediction error, whilst the integration of second sensor further bounds the $R^2$ value within a range close to 1, enforcing accurate predictions. We apply the calculated sensor locations to test data sets. The results for $\operatorname{AOA}_{\max}=6^\circ$ and $\operatorname{AOA}_{\max}=24^\circ$ are shown in Figs. \ref{fig:results2d6} and \ref{fig:results2d24} and compared versus the results without sensors.

\begin{figure}[t]
	\centering
	\includegraphics[]{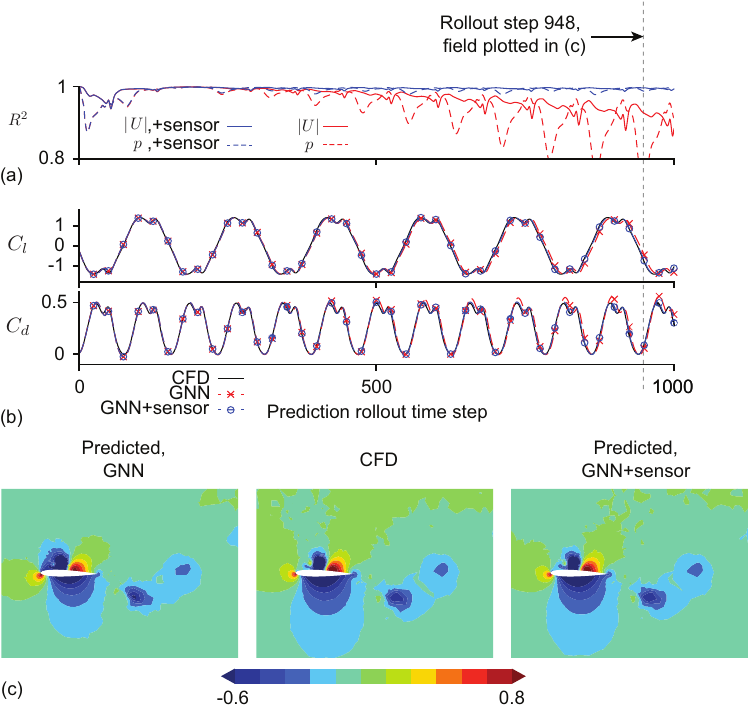}
	\caption{Results for the 2D flow around oscillating airfoil test case at $\operatorname{AOA}_{\max}=26^\circ$. (a) Coefficient of determination $R^2$ of the predicted velocity magnitude and pressure fields over the entire prediction rollout. (b) Lift and drag coefficients calculated from the predicted system states versus the CFD values. (c) Predicted non-dimensionalized pressure field $p^*$ near the airfoil at prediction rollout time step 1000, versus the CFD results.}
	\label{fig:results2d26}
\end{figure}

\begin{figure}[t]
	\centering
	\includegraphics[]{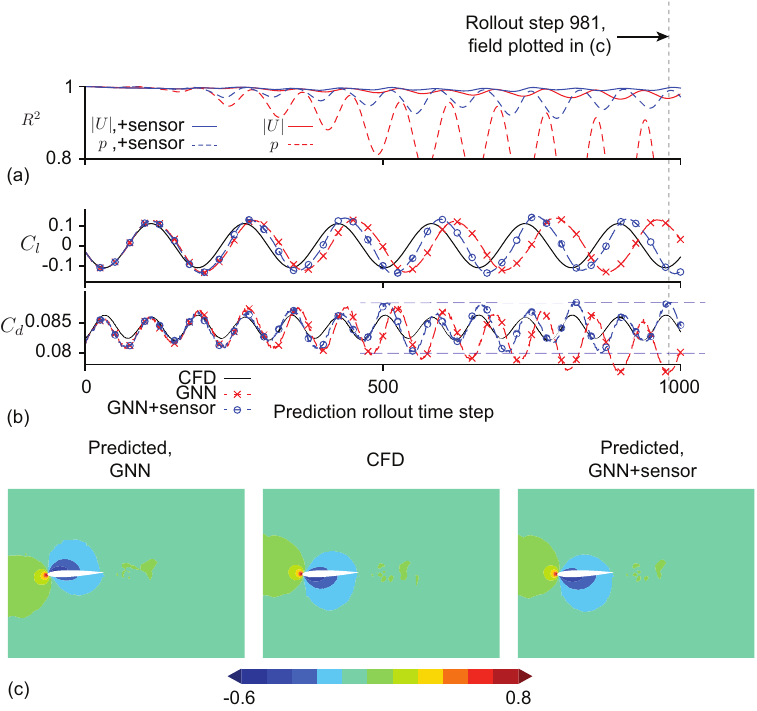}
	\caption{Results for the 2D flow around oscillating airfoil test case at $\operatorname{AOA}_{\max}=2^\circ$. (a) Coefficient of determination $R^2$ of the predicted velocity magnitude and pressure fields over the entire prediction rollout. (b) Lift and drag coefficients calculated from the predicted system states versus the CFD values. (c) Predicted non-dimensionalized pressure field $p^*$ near the airfoil at prediction rollout time step 1000, versus the CFD results.}
	\label{fig:results2d2}
\end{figure}

\begin{figure}
	\centering
	\includegraphics[]{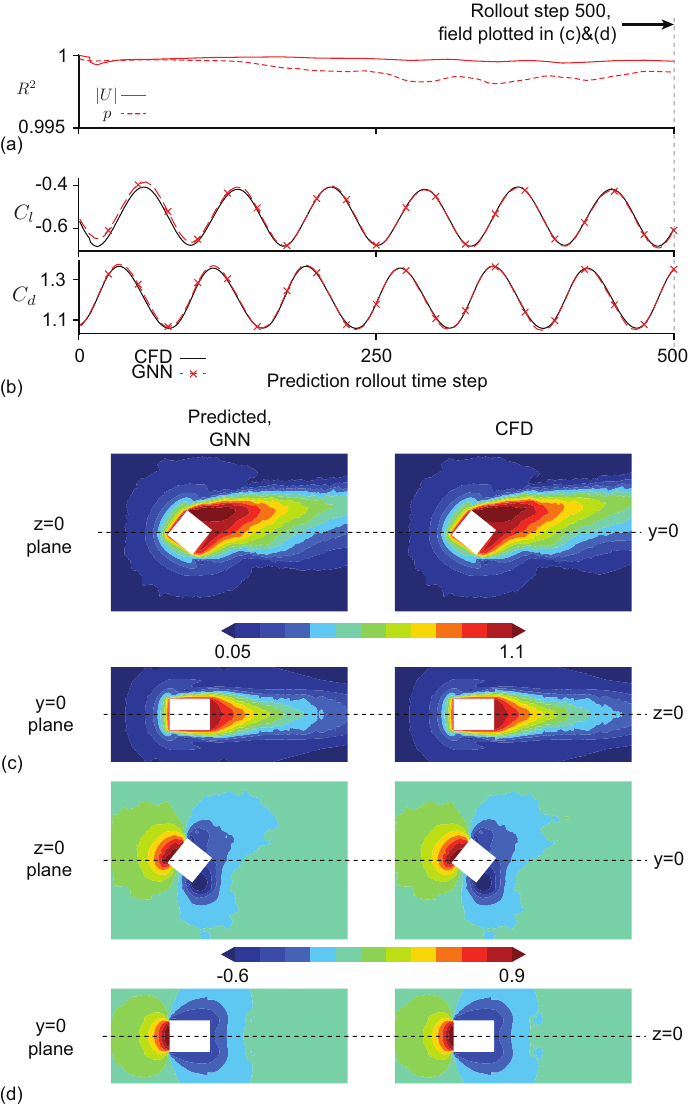}
	\caption{Results for the 3D flow around rotating cube test case. (a) Coefficient of determination $R^2$ for the predicted velocity and pressure fields. (b) Lift and drag coefficients calculated from the predicted states versus the CFD results. (c) The predicted normalized velocity magnitude field after subtracting the inlet velocity $|\boldsymbol{U}^*-(1,0,0)|=|\boldsymbol{U}-(U_\infty,0,0)|/U_\infty$ versus the CFD results at prediction rollout time step 500, and (d) the predicted normalized pressure field $p^*$ versus the CFD results at prediction rollout time step 500.}
	\label{fig:results3d}
\end{figure}

\paragraph*{Behavior in extrapolation range}
Out of the interpolation range, the network is able to make stable predictions in a moderate range, as marked in Fig. \ref{fig:aoaregime}. When the pressure measurements from the two sensors are incorporated, we observe that the network is able to make accurate predictions close to the interpolation range. An example at $\operatorname{AOA}_{\max}=26^\circ$ is reported in Fig. \ref{fig:results2d26}. Further away from the training range, the error becomes more significant, as is shown in Fig. \ref{fig:results2d2}. It is worth pointing out that the whilst the use of sensors cannot help enforce accurate prediction results here, it still enforces a bound on the prediction error, as one could observe from the drag force curve in Fig. \ref{fig:results2d2}b.

\subsection{Experiment -- Extension to 3D}
For the 3D case, we test the trained network on the test flow data set. Similar to that of the 2D case, we start from the first time step within the test data set, and generate roll-out predictions of the system states (velocity and pressure) over the following 500 time steps. Stable and accurate predictions are obtained. We present these results in this subsection.

\paragraph*{System state predictions}
For the 3D test case, we again plot the coefficient of determination $R^2$ for velocity magnitude $\left|\boldsymbol{U}\right|$ and pressure $p$ in Fig. \ref{fig:results3d}a, with the predicted versus ground truth velocity magnitude and pressure fields in sampling planes $y=0$ and $z=0$ plotted in Fig. \ref{fig:results3d}c-\ref{fig:results3d}d. Stable and accurate predictions are obtained for the 500 rollout time steps, as demonstrated by the accuracy of the predicted fields and the consistent $R^2$ value close to 1.

\paragraph*{Lift and drag force predictions}
We plot the predicted versus ground truth drag coefficients in Fig. \ref{fig:results3d}b. One could notice that the lift and drag coefficients calculated from the graph neural network predictions match well with those calculated from the CFD results, showing that the predictions of the system states at the boundary layer are accurate. It is worth noting that the predicted lift coefficients deviate from the CFD values slightly at the starting steps, but self-correct back to the CFD values after about 200 time steps. 

\section{Conclusion}
\label{sec:conclusion}
In this work, we proposed a hypergraph neural network ($\phi$-GNN) framework for modeling unsteady fluid flow around rotating rigid bodies. Separating the domain into co-rotating and static parts, we constructed a single layer of cells at their interface. We allowed these cells to distort and adapt with the rotation of the solid body, and introduced velocity and pressure reconstruction and re-projection schemes to counter the error caused by the the distortion and adaptation of interface cells. 
Operating with mixed floating-point precisions, the proposed framework was applied to two test cases, namely the flow around an oscillating airfoil in 2D and the flow around a rotating cube in 3D. In both test cases, we observed that the framework is able to produce stable predictions for several hundreds or even a thousand time steps. We have further shown that one could establish a digital twin model by incorporating the measurements from sparse sensors, and such a digital twin enforces accurate, error-bounded predictions over the entire interpolation range, with some extrapolation capability. Accurate lift and drag force coefficients on the rotating solid body can be obtained via direct integration of the Cauchy stress tensor on the solid surface. In our future work, we plan to extend the application of the framework to fluid flow around more complex geometries like propellers and turbines.

\bibliography{references}

\begin{thebibliography}{10}
\expandafter\ifx\csname url\endcsname\relax
  \def\url#1{\texttt{#1}}\fi
\expandafter\ifx\csname urlprefix\endcsname\relax\def\urlprefix{URL }\fi
\expandafter\ifx\csname href\endcsname\relax
  \def\href#1#2{#2} \def\path#1{#1}\fi

\bibitem{keck2008thirty}
H.~Keck, M.~Sick, Thirty years of numerical flow simulation in hydraulic
  turbomachines, Acta mechanica 201~(1) (2008) 211--229.

\bibitem{pinto2017computational}
R.~N. Pinto, A.~Afzal, L.~V. D’Souza, Z.~Ansari, A.~Mohammed~Samee,
  Computational fluid dynamics in turbomachinery: a review of state of the art,
  Archives of Computational Methods in Engineering 24~(3) (2017) 467--479.

\bibitem{cheng2022unified}
Z.~Cheng, F.-S. Lien, E.~Yee, H.~Meng, A unified framework for aeroacoustics
  simulation of wind turbines, Renewable Energy 188 (2022) 299--319.

\bibitem{posa2022hydroacoustic}
A.~Posa, R.~Broglia, M.~Felli, M.~Cianferra, V.~Armenio, Hydroacoustic analysis
  of a marine propeller using large-eddy simulation and acoustic analogy,
  Journal of Fluid Mechanics 947 (2022) A46.

\bibitem{li2024deep}
G.~Li, H.~Sun, J.~He, X.~Ding, W.~Zhu, C.~Qin, X.~Zhang, X.~Zhou, B.~Yang,
  Y.~Guo, Deep learning, numerical, and experimental methods to reveal
  hydrodynamics performance and cavitation development in centrifugal pump,
  Expert Systems with Applications 237 (2024) 121604.

\bibitem{li2024fast}
C.~Li, B.~Liang, P.~Yuan, Q.~Zhang, Y.~Liu, B.~Liu, M.~Zhao, Fast prediction of
  propeller dynamic wake based on deep learning, Physics of Fluids 36~(8)
  (2024).

\bibitem{hou2024reconstruction}
X.~Hou, X.~Zhou, Y.~Liu, Reconstruction of ship propeller wake field based on
  self-adaptive loss balanced physics-informed neural networks, Ocean
  Engineering 309 (2024) 118341.

\bibitem{gao2022simulation}
F.~Gao, Z.~Zhang, C.~Jia, Y.~Zhu, C.~Zhou, J.~Wang, Simulation and prediction
  of three-dimensional rotating flows based on convolutional neural networks,
  Physics of Fluids 34~(9) (2022).

\bibitem{zhang2024rotating}
M.~Zhang, L.~Wang, Y.-B. Xu, X.-L. Wang, Rotating machinery flow field
  prediction based on hybrid neural network, Journal of Turbulence 25~(12)
  (2024) 482--500.

\bibitem{liu2024parameterized}
K.~Liu, K.~Luo, Y.~Cheng, A.~Liu, H.~Li, J.~Fan, S.~Balachandar, Parameterized
  physics-informed neural networks (p-pinns) solution of uniform flow over an
  arbitrarily spinning spherical particle, International Journal of Multiphase
  Flow 180 (2024) 104937.

\bibitem{zhang2024performance}
M.~Zhang, C.~Wang, J.~Zhang, Performance prediction of co-rotating disk cavity
  with finned vortex reducer based on machine learning, International Journal
  of Thermal Sciences 205 (2024) 109287.

\bibitem{pfaff2020learning}
T.~Pfaff, M.~Fortunato, A.~Sanchez-Gonzalez, P.~W. Battaglia, Learning
  mesh-based simulation with graph networks, arXiv preprint arXiv:2010.03409
  (2020).

\bibitem{lino2022multi}
M.~Lino, S.~Fotiadis, A.~A. Bharath, C.~D. Cantwell, Multi-scale
  rotation-equivariant graph neural networks for unsteady eulerian fluid
  dynamics, Physics of Fluids 34~(8) (2022) 087110.

\bibitem{lino2024se}
M.~Lino, N.~Thuerey, T.~Pfaff, Se (3)-equivariant diffusion graph nets:
  Synthesizing flow fields by denoising invariant latents on graphs, in: ICML
  2024 AI for Science Workshop, 2024.

\bibitem{fortunato2022multiscale}
M.~Fortunato, T.~Pfaff, P.~Wirnsberger, A.~Pritzel, P.~Battaglia, Multiscale
  meshgraphnets, arXiv preprint arXiv:2210.00612 (2022).

\bibitem{cao2022bi}
Y.~Cao, M.~Chai, M.~Li, C.~Jiang, Bi-stride multi-scale graph neural network
  for mesh-based physical simulation, arXiv preprint arXiv:2210.02573 (2022).

\bibitem{gao2024finite}
R.~Gao, I.~K. Deo, R.~K. Jaiman, A finite element-inspired hypergraph neural
  network: Application to fluid dynamics simulations, Journal of Computational
  Physics (2024) 112866.

\bibitem{gao2024predicting}
R.~Gao, R.~K. Jaiman, Predicting fluid--structure interaction with graph neural
  networks, Physics of Fluids 36~(1) (2024).

\bibitem{ma2022fast}
Z.~Ma, Z.~Ye, W.~Pan, Fast simulation of particulate suspensions enabled by
  graph neural network, Computer Methods in Applied Mechanics and Engineering
  400 (2022) 115496.

\bibitem{xu2021conditionally}
J.~Xu, A.~Pradhan, K.~Duraisamy, Conditionally parameterized,
  discretization-aware neural networks for mesh-based modeling of physical
  systems, Advances in Neural Information Processing Systems 34 (2021)
  1634--1645.

\bibitem{gao2024towards}
R.~Gao, S.~Heydari, R.~K. Jaiman, Towards spatio-temporal prediction of
  cavitating fluid flow with graph neural networks, International Journal of
  Multiphase Flow 177 (2024) 104858.

\bibitem{li2024physics}
S.~Li, Z.~Sun, Y.~Zhu, C.~Zhang, Physics-constrained and
  flow-field-message-informed graph neural network for solving unsteady
  compressible flows, Physics of Fluids 36~(4) (2024).

\bibitem{kazadi2024floodgnn}
A.~Kazadi, J.~Doss-Gollin, A.~Sebastian, A.~Silva, Floodgnn-gru: a
  spatio-temporal graph neural network for flood prediction, Environmental Data
  Science 3 (2024) e21.

\bibitem{tang2024graph}
H.~Tang, L.~J. Durlofsky, Graph network surrogate model for subsurface flow
  optimization, Journal of Computational Physics 512 (2024) 113132.

\bibitem{zhao2024review}
Y.~Zhao, H.~Li, H.~Zhou, H.~R. Attar, T.~Pfaff, N.~Li, A review of graph neural
  network applications in mechanics-related domains, Artificial Intelligence
  Review 57~(11) (2024) 315.

\bibitem{li2023uncertainty}
J.~Li, T.~Liu, G.~Zhu, Y.~Li, Y.~Xie, Uncertainty quantification and
  aerodynamic robust optimization of turbomachinery based on graph learning
  methods, Energy 273 (2023) 127289.

\bibitem{murphy1994cfd}
J.~Murphy, Cfd simulation of flows in stirred tank reactors using a sliding
  mesh technique, in: Instn. Chem. Engng. Symp. Ser., Vol. 136, 1994, pp.
  341--348.

\bibitem{behr1999shear}
M.~Behr, T.~Tezduyar, The shear-slip mesh update method, Computer Methods in
  Applied Mechanics and Engineering 174 (1999) 261--274.

\bibitem{weller1998tensorial}
H.~G. Weller, G.~Tabor, H.~Jasak, C.~Fureby, A tensorial approach to
  computational continuum mechanics using object-oriented techniques, Computers
  in physics 12~(6) (1998) 620--631.

\bibitem{issa1986solution}
R.~I. Issa, Solution of the implicitly discretised fluid flow equations by
  operator-splitting, Journal of computational physics 62~(1) (1986) 40--65.

\bibitem{holzer2008new}
A.~H{\"o}lzer, M.~Sommerfeld, New simple correlation formula for the drag
  coefficient of non-spherical particles, Powder Technology 184~(3) (2008)
  361--365.

\bibitem{haider1989drag}
A.~Haider, O.~Levenspiel, Drag coefficient and terminal velocity of spherical
  and nonspherical particles, Powder technology 58~(1) (1989) 63--70.

\bibitem{augier2019fluiddyn}
P.~Augier, A.~V. Mohanan, C.~Bonamy, {FluidDyn}: A python open-source framework
  for research and teaching in fluid dynamics by simulations, experiments and
  data processing, Journal of Open Research Software 7 (2019).

\bibitem{virtanen2020scipy}
P.~Virtanen, R.~Gommers, T.~E. Oliphant, M.~Haberland, T.~Reddy, D.~Cournapeau,
  E.~Burovski, P.~Peterson, W.~Weckesser, J.~Bright, S.~J. {van der Walt},
  M.~Brett, J.~Wilson, K.~J. Millman, N.~Mayorov, A.~R.~J. Nelson, E.~Jones,
  R.~Kern, E.~Larson, C.~J. Carey, {\.I}.~Polat, Y.~Feng, E.~W. Moore,
  J.~{VanderPlas}, D.~Laxalde, J.~Perktold, R.~Cimrman, I.~Henriksen, E.~A.
  Quintero, C.~R. Harris, A.~M. Archibald, A.~H. Ribeiro, F.~Pedregosa, P.~{van
  Mulbregt}, {SciPy 1.0 Contributors}, {{SciPy} 1.0: Fundamental Algorithms for
  Scientific Computing in Python}, Nature Methods 17 (2020) 261--272.

\bibitem{paszke2019pytorch}
A.~Paszke, S.~Gross, F.~Massa, A.~Lerer, J.~Bradbury, G.~Chanan, T.~Killeen,
  Z.~Lin, N.~Gimelshein, L.~Antiga, et~al., Pytorch: An imperative style,
  high-performance deep learning library, Advances in neural information
  processing systems 32 (2019).

\bibitem{sitzmann2020implicit}
V.~Sitzmann, J.~Martel, A.~Bergman, D.~Lindell, G.~Wetzstein, Implicit neural
  representations with periodic activation functions, Advances in Neural
  Information Processing Systems 33 (2020) 7462--7473.

\bibitem{kingma2014adam}
D.~P. Kingma, J.~Ba, Adam: A method for stochastic optimization, arXiv preprint
  arXiv:1412.6980 (2014).

\bibitem{micikevicius2018mixed}
P.~Micikevicius, S.~Narang, J.~Alben, G.~Diamos, E.~Elsen, D.~Garcia,
  B.~Ginsburg, M.~Houston, O.~Kuchaiev, G.~Venkatesh, et~al., Mixed precision
  training, in: International Conference on Learning Representations, 2018.

\bibitem{list2024differentiability}
B.~List, L.-W. Chen, K.~Bali, N.~Thuerey, Differentiability in unrolled
  training of neural physics simulators on transient dynamics, Computer Methods
  in Applied Mechanics and Engineering 433 (2025) 117441.

\bibitem{sanchez2020learning}
A.~Sanchez-Gonzalez, J.~Godwin, T.~Pfaff, R.~Ying, J.~Leskovec, P.~W.
  Battaglia, Learning to simulate complex physics with graph networks, in: 37th
  International Conference on Machine Learning, 2020.

\end{thebibliography}

\end{document}